# Strongly anisotropic antiferromagnetic coupling in EuFe$_2$As$_2$ revealed by stress detwinning

**Authors:** Joshua J Sanchez[1], Gilberto Fabbris[3], Yongseong Choi[3], Yue Shi[1], Paul Malinowski[1], Shashi Pandey[2], Jian Liu[2], I.I. Mazin[5], Jong-Woo Kim[3], Philip Ryan[3,4], Jiun-Haw Chu[1*]

**Affiliation:**

[1] Department of Physics, University of Washington, Seattle, Washington 98195, USA.

[2] Department of Physics and Astronomy, University of Tennessee, Knoxville, Tennessee 37996, USA.

[3] Advanced Photon Source, Argonne National Laboratories, Lemont, Illinois 60439, USA.

[4] School of Physical Sciences, Dublin City University, Dublin 9, Ireland.

[5] Department of Physics and Astronomy and Quantum Science and Engineering Center, George Mason University, Fairfax, VA 22030

*Correspondence to: jhchu@uw.edu (J.-H.C)

**Abstract:**

Of all parent compounds of iron-based high-temperature superconductors, $EuFe_2As_2$ exhibits by far the largest magnetostructural coupling due to the sizable biquadratic interaction between Eu and Fe moments. While the coupling between Eu antiferromagnetic order and Fe structural/antiferromagnetic domains enables rapid field detwinning, this prevents simple magnetometry measurements from extracting the critical fields of the Eu metamagnetic transition. Here we measure these critical fields by combining x-ray magnetic circular dichroism spectroscopy with in-situ tunable uniaxial stress and applied magnetic field. The combination of two tuning knobs allows us to separate the stress-detwinning of structural domains from the field-induced reorientation of Eu moments. Intriguingly, we find a spin-flip transition which can only result from a strongly anisotropic interaction between Eu planes. We argue that this anisotropic exchange is a consequence of the strong anisotropy in the magnetically ordered Fe layer, which presents a new form of higher-order coupling between Eu and Fe magnetism.

Magnetism is the origin of a wide range of intriguing phenomena in iron-based superconductors, including electronic nematicity and high temperature superconductivity[1]–[5]. In contrast to the magnetism of the high-$T_c$ cuprates, the stripe spin density wave (SDW) ground state breaks fourfold rotational symmetry and the spin dynamics are highly anisotropic [6]–[9]. Key to this highly anisotropic magnetism is a sizable biquadratic coupling that is not captured in a simple Heisenberg model and likely arises from the dual itinerant-localized character of the Fe magnetism[10], [11]. It has also been suggested that this biquadratic term is necessary for the emergence of spin nematicity [12], [13].

Very recently, it was shown that a similar biquadratic coupling plays a role in the unusually large magnetoelastic coupling in EuFe$_2$As$_2$, where the introduction of a magnetic lanthanide element adds another layer of complexity into the magnetism of iron-based superconductors[14]. In addition to the SDW order, EuFe$_2$As$_2$ also hosts an A-type layered antiferromagnetic (AFM) order in the Eu layer, sharing the same easy axis with the Fe-SDW. Like in other iron pnictides, the SDW in EuFe$_2$As$_2$ creates orthorhombic twin domains. In BaFe$_2$As$_2$, due to the strong coupling between the structural distortion and the SDW, an applied field of order 25 T is able to fully detwin the structural domains [15]. Surprisingly, EuFe$_2$As$_2$ can be fully detwinned with less than 1 T, and partial detwinning can persist even after the field is turned off [16], [17]. Magnetization, NMR and neutron diffraction data show that this structural detwinning coincides with the reorientation of Eu moments towards the applied field direction, suggesting that the Eu magnetism and the associated large magnetic moments are responsible for this drastic reduction of detwinning field [16], [18], [19]. Nevertheless, as no single-ion anisotropy is present for the half-filled Eu $4f^7$ electrons and no dipolar coupling between the Fe and Eu layers is allowed by symmetry, it remained an open question how Eu moments even sense the orthorhombic direction. Recently, Maiwald et al solved this mystery by considering a biquadratic coupling between Fe-SDW and Eu-AFM moments[14]. The biquadratic coupling of the form $K(f_i \cdot e_j)^2$, where $f_i$ and $e_j$ represent the Fe and Eu moments, respectively, provides an effective single-ion anisotropy that couples the Eu moment orientation with the Fe-SDW direction[14]. Therefore, while the Fe-Fe biquadratic coupling generates the nematicity in the FeAs layer, the Eu-Fe biquadratic coupling provides a pathway for the Eu magnetism to couple to the structural orthorhombicity and the underlying nematicity.

Here, we report the discovery of another consequence of the higher order Eu-Fe coupling in EuFe$_2$As$_2$ – a highly anisotropic Eu-Eu interplanar coupling. The degree of anisotropy of Eu-Eu interlayer exchange ($\frac{J_x - J_y}{J_x + J_y}$) is about 75 times larger than the structural orthorhombicity, which can only be understood by considering the influence of the Fe-SDW order. The anisotropy of Eu-Eu interlayer exchange was overlooked previously because the field detwinning process masked the spin-flip nature of Eu metamagnetic transition. We overcome this challenge by a direct measurement of the Eu metamagnetic transition in a mechanically detwinned sample using a piezoelectric stress device, which allows us to apply magnetic fields either parallel or perpendicular to the easy axis of Eu moments within a single structural domain. Conventional magnetometry techniques are difficult to apply to a sample mounted to a strain device due to the added size and background magnetization contributed by the device. We therefore employ x-ray magnetic circular dichroism (XMCD) on the Eu L$_3$ edge to measure the Eu-specific in-plane magnetization induced by an applied magnetic field. We show that we are able to strain the crystal into a monodomain which exhibits either a large jump in magnetization, consistent with a spin flip transition, or has a perfectly linear magnetization from continuous canting of Eu moments (i.e., we are able to turn the metamagnetic transition on and off). From the measurement of the critical field for the Eu spin flip and the field dependence of the spin canting, we determine the energies of the Eu-Eu and Eu-Fe coupling and discover that the Eu-Eu interaction itself is directionally dependent on the orientation of Fe moments. We then confirm this by first principles calculations. The discovery of the anisotropic Eu-Eu interplanar coupling also sheds light on the evolution of the Eu magnetism in doped EuFe$_2$As$_2$ which we reevaluate in the discussion.

**RESULTS:**

Figure 1a shows the fully magnetically ordered unit cell at T=7K and zero applied magnetic field. Eu moments are aligned ferromagnetically within each plane and antiferromagnetically between planes. The Fe spin density wave ordering results in a small structural orthorhombicity and the formation of structural twin domains that are identical up to a 90-degree rotation. Within each domain, the Eu AFM and Fe SDW easy axes are aligned with the longer orthorhombic $a$ lattice constant [20], [21]. We use a geometry with the $\hat{x}$ and $\hat{y}$ axes aligned to the orthorhombic $a$ and $b$ directions, and tensile or compressive stress applied along the $\hat{x}$ direction detwins the sample toward the A monodomain ($a$ lattice vector along $\hat{x}$) or B monodomain ($b$ lattice vector along $\hat{x}$), respectively (Fig.1b). Given the maximum orthorhombicity of EuFe$_2$As$_2$ (0.28% at 2.5K [21]) and the maximum strain of the sample device at 7K (0.3%), we are capable of nearly fully detwinning the sample (the minor domain is estimated to be 5% or less of the sample volume, see Supplementary Figure 2 and ref. [22] ). Once in the A (B) monodomain state, the measured resistivity $\rho_{xx}$ becomes sensitive to the anisotropic resistivity $\rho_a$ ($\rho_b$) along the orthorhombic $a$ ($b$) direction. Magnetic field is applied perpendicular to the current/strain axis and at 10 degrees above the ab plane. Except for a change in the strain state, the sample is not reoriented in any way during the experiment, ensuring identical effective fields and XMCD-illuminated sample volumes. XMCD measures the induced Eu magnetic moment along the field direction, which is fixed parallel to the incident x-ray direction in this study.

First, we address the effect of the applied magnetic field at T=30K, below the orthorhombic and SDW transitions ($T_{SDW} = 187K$), but above the Eu AFM ordering temperature ($T_N = 19.1K$). In the Eu$^{2+}$ valence state, the $4f^7$ electrons have zero orbital angular momentum ($L = 0$), and as such are expected to show an isotropic response to applied field. After detwinning to either the A or B monodomain, we applied fields from 0 to 1 T and measured the XMCD and resistivity simultaneously in 0.02 T steps (XMCD data were not collected for the B monodomain for 0.8 T-1 T at this temperature). We find an XMCD signal that is indistinguishable between the two domains, suggesting that at this temperature and field range the Eu-Fe interaction is negligible compared to the Eu paramagnetic coupling to applied field (Fig.2). We note that compared to the data presented next, the XMCD values at 1 T and 30 K are roughly 3 times smaller than the 1 T XMCD saturation value within the Eu AFM phase, consistent with a lower susceptibility in the paramagnetic phase. As in BaFe$_2$As$_2$, the zero-field resistivity is considerably larger along the orthorhombic $b$ direction than the $a$ direction. Both $\rho_a$ and $\rho_b$ have a weak field dependence at 30K. The inset to Fig. 2 shows the detwinned sample cooling through the Eu AFM transition. As demonstrated previously in mechanically detwinned EuFe$_2$As$_2$, we see no change in the Eu transition temperature between the tensile and compressive cooling data, nor do we see any additional resistivity anisotropy induced by the Eu AFM ordering [23]. Indeed, the resistivity anisotropy $\frac{\rho_b - \rho_a}{\rho_b + \rho_a} = 0.084(2)$ is unchanged through this temperature range.

We now discuss results from within the Eu AFM phase at 7K. We applied field through a $\pm 1$ T loop to each detwinned state and measured the XMCD and resistivity simultaneously in 0.02 T steps (Fig.3). In the A domain (Fig.3a), the Eu moments are initially aligned with the AFM easy axis transverse to the field direction. The linear growth of XMCD signal with field indicates that Eu moments cant continuously to align with the field, with no observable hysteresis. Conversely, in the B domain (Fig.3b) the easy axis is along the field direction, and so for fields below 0.4 T the XMCD is nearly flat as no canting can occur. The jump in XMCD from 0.4 T to 0.6 T and accompanying hysteresis is a clear sign of a metamagnetic spin flip transition between the Eu AFM and FM states. The weak linear field dependence of magnetization in the pre- and post-spin-flip field ranges are due to the out of plane magnetization induced by the small out of plane field component, which is assumed to equally contribute to the A domain magnetization. We note that the observation of spin flip transition contradicts the expectation from the spin Hamiltonian derived from ref[14], which predicts a spin flop transition. As will be shown in the discussion section, this contradiction can only be resolved by including an anisotropic interlayer exchange between Eu moments.

The resistivity is approximately linear in field for $|\mu_0 H| < 0.4T$ and $|\mu_0 H| > 0.6T$ for both domains. For $0.4T < |\mu_0 H| < 0.6T$, a large hysteretic drop occurs in the B domain resistivity, coinciding with the jump in XMCD. A much smaller drop also occurs in the A domain, which also shows a small hysteresis (Fig.3a, inset), and is likely due to a remnant B domain that was not fully detwinned and which is not resolved in the XMCD measurement. For each domain, the resistivity returns to the initial zero-field value after the field loop, indicating that there is no persistent field-detwinning in our setup as is found reliably from previous studies of freestanding samples [16], [17]. Further, given the field and current orientation, the sample in either strain state would be expected to detwin towards the higher-resistivity B domain above ~0.5T, and so the drop in resistivity for both domains suggests that our stress device is indeed preventing field detwinning. This is also strong evidence for a purely Eu spin origin of the resistivity jump.

Both the XMCD and resistivity data show excellent agreement between positive and negative field values. In Figure 4a, we plot the average value over positive and negative field sweeps of the XMCD and the magnetoresistance. While we were unable to apply enough field to fully saturate the A domain XMCD, we can extrapolate the field dependencies of each domain to estimate the saturation field. Linear fits to the XMCD of the A and B domains at fields greater than 0.6 T are shown, with an intersection at $\mu_0 H_A^{sat}$ = 1.17 T. Beyond this field, the Eu magnetic moment is expected to be fully saturated in each domain, as seen in freestanding crystal magnetometry studies [14], [17], [18], [24]. To more precisely determine the B domain spin flip field $H_B^{flip}$ we use the value at the center of the field hysteresis. Figure 4b shows the difference between the increasing and decreasing field values of the XMCD and resistivity. In the B domain, a sharp peak in both quantities occurs at $\mu_0 H_B^{flip} = 0.48T$. In principle, the lattice distortions induced by the detwinning strain could cause a change in the interaction strengths between Eu and Fe planes that could alter these critical fields. In Supplementary Figure 3 we present data for the same sample tuned to a nearly-zero strain state in which both A and B domains are present and show that $H_A^{sat}$ and $H_B^{flip}$ are essentially unchanged, demonstrating that the strain applied to detwin the sample does not appreciably affect the interplanar coupling strengths.

**DISCUSSION**

We can relate the spin flip and saturation fields $H_B^{flip}$ and $H_A^{sat}$ to the microscopic interactions in the sample using a spin Hamiltonian (Supplementary Information section IV). We will start with a modified version of the spin Hamiltonian presented in ref. [14] and show that to explain the experimental results an additional anisotropic exchange term is needed. We consider first a purely isotropic exchange energy $2J$ between Eu antiferromagnetic planes in the doubled unit cell of the fully ordered state ($\text{Eu}_2\text{Fe}_4\text{As}_4$). From the symmetry of the magnetic ordering structure (Fig. 1a), the dipolar interactions of Eu and Fe moments cancel each other and so do not contribute to the magnetic energy. The half-filled $4f^7$ orbital of $\text{Eu}^{2+}$ has zero orbital angular momentum ($L = 0$) and negligible single ion anisotropy, but the posited biquadratic coupling between Eu and Fe moments creates an effective magnetocrystalline anisotropy. We define a biquadratic coupling energy $K$ between the 8 inequivalent Eu-Fe moment pairs, with a total Eu-Fe planar coupling energy $8K$. Following the usual treatment of spin-flip and spin canting transitions we determine the critical fields as

$$H_B^{flip} = \frac{2J}{M}$$

$$H_A^{sat} = \frac{4J + 16K}{M} = 2\left(H_B^{flip} + \frac{8K}{M}\right)$$

For the B domain, Eu and Fe moments are aligned both before and after the spin flip, and so the presence of the biquadratic coupling does not change the value of the spin flip field, but only serves to provide the necessary magnetocrystalline anisotropy to enforce a sharp spin flip transition. For the A domain, the biquadratic coupling provides an extra energy barrier that must be overcome by the field to reach the fully saturated canted state. From the measured values $\mu_0 H_A^{sat} = 1.17$ T and $\mu_0 H_B^{flip} = 0.48$ T, and using the expected Eu moment $M = 6.8\ \mu_B$ [16], [17], we obtain $J = 94.5\ \mu eV$ and $8K = 41.3\ \mu eV$. Thus, the Eu-Eu and Eu-Fe planar coupling energies are comparable, $\frac{J}{8K} = 2.3$.

At this point we are facing an apparent contradiction: we observe experimentally a sharp spin *flip* transition, but for $\frac{J}{8K} > 1$ the spin Hamiltonian would actually be expected to result in a spin *flop* transition, as shown in Fig. 5(a). ( See ref. [14] and Supplementary Information section IV for further discussion). This discrepancy has important implications for correctly modelling the field detwinning process, which has been based on the (never actually observed) spin flop transition. This apparent contradiction indicates that an additional term is needed in the spin Hamiltonian. The simplest such term is a symmetric anisotropic exchange term $W(e_{i,x}e_{i+1,x} - e_{i,y}e_{i+1,y})$, where $e_{i,x}$ and $e_{i,y}$ are the $x$ and $y$ components of the Eu moment in the $i$ th layer, using a notation with Fe moments aligned antiferromagnetically along $\hat{x}$. This term then increases (decreases) the interaction strength between Eu moments when aligned parallel (perpendicular) to the Fe moments, and provides the additional anisotropy needed to enforce a spin flip transition. Although other higher order terms could also be introduced to the spin Hamiltonian, they generally lead to non-linear *M-H* curves which were not observed in the experiment. With this additional term, the criterion for a spin flip transition becomes $\frac{J}{8K+W} < 1$, and the critical fields are:

$$H_B^{flip} = \frac{2(J+W)}{M}$$

$$H_A^{sat} = \frac{4J + 16K}{M} = 2\left(H_B^{flip} + \frac{8K - 2W}{M}\right).$$

The three parameters $J$, $K$ and $W$ cannot be uniquely determined by the two experimentally measured values. Therefore, additional constraint is needed. This constraint can be provided by the measurement of $H_{45°}^{sat}$, i.e. the field required to saturate the magnetization when the field is aligned 45 degrees to the easy and hard axis, such that the field has an equivalent effect on both domains. It has the following expression:

$$H_{45°}^{sat} = \frac{4J}{M}.$$

As our strain+XMCD measurements have demonstrated the need for the $W$ term, we now discuss the extraction of $J$, $K$ and $W$ from magnetization measurements using a vibrating sample magnetometer (VSM) on a second sample from the same growth batch. We stress that this extraction is not possible without the confirmation of a spin-flip transition by XMCD measurement on a stress-detwinned sample. The sample was encased in GE varnish so that the domain configuration is fixed to 50/50 and the field detwinning is prohibited. The thin octagonal shape of the sample also ensures that field applied totally in-plane along the $[1\ 0\ 0]_T$ and $[1\ 1\ 0]_T$ directions have an identical (negligible) demagnetization factor (see Methods). As shown in Figure 5b, at T=2K for field applied along the $[1\ 0\ 0]_T$ direction (blue curve), i.e. at 45° to both domain easy axes, the magnetization exhibits continuous spin canting toward saturation at $\mu_0 H_{45°}^{sat} = 0.73$ T. For field applied along the $[1\ 1\ 0]_T$ direction (red curve) the *M* vs *H* perfectly overlaps with the combination of the responses of A and B domains from the XMCD measurements (grey), confirming the absence of field detwinning. We extract $\mu_0 H_A^{sat} = 1.15$ T and $\mu_0 H_B^{flip} = 0.45$ T from these spin flip and saturation fields, which is in good agreement with the XMCD-measured values. Using these 3 measured values, we can uniquely solve for the interaction terms and find $J = 71.8\ \mu eV$, W= 16.7 $\mu eV$ and 8K = 82.7 $\mu eV$, with a ratio $\frac{J}{8K+W} = 0.72 < 1$ satisfying the sharp spin flip criterion. Therefore, we find that the Eu-Eu interaction is much stronger for Eu moments aligned parallel ($J_x = J + W = 88.5\ \mu eV$) compared to perpendicular ($J_y = J - W = 55.1\ \mu eV$) to Fe moments. Further, the normalized difference of the anisotropic interplanar interaction, $\frac{J_x-J_y}{J_x+J_y} = 23.3\%$, is nearly two orders of magnitude greater than the corresponding normalized difference of in-plane lattice constants (the orthorhombicity), $\frac{a-b}{a+b} \sim 0.28\%$, which strongly implies the Fe-SDW origin of the anisotropy.

To gain more insight, we used density functional theory (DFT) to calculate the exchange coupling between Eu layers as the difference between ferro- and antiferromagnetically stacked Eu layers, for Eu moments parallel and perpendicular to the Fe moments. We used the standard VASP package [25], [26], and verified that the results were fully converged with respect to the Brillouin zone integration, plane wave cutoff, and the number of bands included in the diagonalization. We also varied the effective Hubbard repulsion parameter for Eu $f$ orbitals, U-J, between 5 and 7 eV, which had little impact on the result. We find that for parallel Eu-Fe moments, $J_x$=150-170 $\mu eV$, while for perpendicular Eu-Fe moments, $J_y$ is essentially zero. While this result seems to considerably overestimate the exchange anisotropy (see ref. [14], Supplementary Materials Section 1A, for a discussion of the difficulties of DFT calculation for noncollinear Eu-Fe moments, which we assume contributes to this overestimation), it clearly shows that DFT calculations also support a strongly anisotropic Eu interplanar interaction.

The anisotropic interaction between Eu planes can be thought to result from the anisotropic hopping of conduction electrons through the Fe plane. Considering a standard superexchange interaction, conduction electrons with spins polarized along the Eu direction will generally have a larger Eu-Fe hopping amplitude $t$ when the Eu and Fe moments are parallel ($J_x = J + W$) rather than perpendicular ($J_y = J - W$), which generates a stronger antiferromagnetic interaction for collinear Eu-Fe moments. Further, the weak but finite *ferromagnetic* interaction between Eu planes is also mediated through the Fe layer, and as the Fe moments have a much larger susceptibility perpendicular to their ordering direction, the ferromagnetic interaction is stronger for perpendicular Eu and Fe moments, which weakens their overall effective *antiferromagnetic* interaction (this can be considered an extreme case of the RKKY interaction).

In this sense, we can view this Eu interplanar interaction anisotropy (W) as a generalized Eu-Fe biquadratic coupling independent of the previously investigated Eu-Fe biquadratic coupling (K), where the square of Fe moments $\left(f_{i,x}^2 - f_{i,y}^2\right)$ couples the Eu moments above and below the iron plane. We note that this picture is reminiscent of the Fe-Fe biquadratic coupling within the FeAs plane that generates an effective anisotropic in-plane exchange of Fe moments $J_{1a} - J_{1b}$ [11]. These types of interactions have been long overlooked in the past, but are relevant to rare-earth and transition metal intermetallic systems with multiple magnetic orders[27], [28].

A complete determination of the spin Hamiltonian in EuFe$_2$As$_2$ also sheds light on the doping dependence of Eu magnetic order, which has yet to be fully understood. As in other iron pnictides, chemical doping in EuFe$_2$As$_2$ rapidly suppresses the Fe-SDW and stabilizes superconductivity [29]–[37]. In contrast, the doping has only a weak effect on the magnetic ordering temperature in the Eu layer, but causes a smooth evolution from an A-type AFM order to a c-axis canted AFM order (c-AFM), and finally to a c-axis ferromagnetic order (FM) [20], [21], [36]–[42]. This doping dependence can be naturally understood as the consequence of Eu moments lowering their energy by aligning with the Fe-SDW, with this energy saving being gradually diminished as doping weakens the Fe-SDW. Further, in the parent compound both the Fe-SDW and Eu-AFM are robust under moderate hydrostatic pressure even as superconductivity develops [43], while in the underdoped case a pressure-induced transition from c-AFM to FM occurs only after the SDW is nearly fully suppressed[44]. This suggests the SDW plays a role in both the orientation and the interaction of Eu moments. Future doping dependence studies may provide more insight on how the Eu interlayer interaction is influenced by the various orders in the FeAs plane, including by superconductivity[45].

In conclusion, our sample environment allows us to approach the magnetic coupling and magnetotransport properties of the EuFe$_2$As$_2$ system in an unprecedented fashion. Through mechanical stress we can prevent field detwinning and gain access to the meta-magnetism and the associated magnetotransport behavior of a monodomain sample. From measurements of the spin flip and moment saturation fields we are able to determine the strengths of coupling between Eu and Fe planes and discover the presence of an anisotropic exchange term in the spin Hamiltonian. We emphasize again that in a freestanding crystal, the rapid field-detwinning has prevented any previous determination of the anisotropic Eu interplanar interaction in this system. This new technique not only deepens our understanding of the $EuFe_2As_2$ system, but can also be applied to a variety of systems to disentangle the strongly coupled spin, orbit and lattice degree of freedom.

**METHODS:**

**Sample Preparation**

Single crystal samples of $EuFe_2As_2$ were grown from a tin flux as described elsewhere [38]. The sample was cleaved from a large as-grown single crystal plate and cut along the tetragonal [1 1 0] direction into a bar with dimensions 3.2 x 0.50 x 0.065 mm. These sample dimensions correspond to a demagnetization factor of N=0.13 along the applied field direction [46], resulting in a small maximum demagnetization field of only $NM_{Eu}{\sim}0.005T$. A piezo-actuator uniaxial stress device (*Razorbill Instruments, CS*-100) was used to provide in-situ stress in the $B_{2g}$ symmetry channel, i.e. in the direction of the orthorhombic distortion, such that the applied tensile or compressive stress detwins the sample to either the A or B domain, respectively (Fig.1). The four-wire electrical contact geometry is illustrated in Figure 1, with wires underneath the sample to not obstruct the x-ray fluorescence off the top surface of the crystal. Measurements of the resistivity coefficient $\rho_{xx}$ aligned along the stress axis were performed using a standard 4-point measurement and an SR830 lock-in amplifier. A second sample from the same growth batch was cut into a thin octagon with surface area 2.18 mm$^2$ and thickness 0.0165 mm, with negligible in-plane demagnetization factor [46]. Magnetization was measured in a Quantum Design PPMS.

**X-ray Magnetic Circular Dichroism**

XMCD was measured at the Advanced Photon Source beamline 4-ID-D at Argonne National Laboratory. We probed the Eu L$_3$ edge using x-rays of 6.97 keV, which measure the spin polarization of the Eu 5d band due to the magnetic moment of the 4f orbital. Generally, the Eu L$_3$ edge XMCD signal can be taken as proportional to the 4f moment magnetization; however, as the Eu 5d band has a significant hybridization with the As 4p orbitals [47], which themselves hybridize with the Fe 3d orbitals, the exact value of the XMCD is expected to have some dependence on Fe conduction effects. Nonetheless, we can use the sharp changes in XMCD signal to mark the fields at which magnetic transitions and saturations occur at. A superconducting split coil magnet with a large bore was used to apply magnetic field. The sample temperature was controlled using He flow. XMCD was collected in fluorescence geometry by monitoring the Eu L$_\alpha$ line using a four element Vortex detector integrated with the Xspress module to enable a larger dynamical range. Circularly polarized x-rays were generated using a 180 microns thick diamond (111) phase plate. Data was corrected for self-absorption.

The XMCD spot size illuminates the whole sample width across the y direction and is roughly 100 microns wide along the x direction (between the transport wires) and probes a depth of about 5 microns. The beam is centered on the middle of the crystal where strain is most transmitted and homogenous. The transport wires are separated by about 1700 microns, and transport is sensitive to the whole bulk of the sample. While the resistivity and XMCD are not measuring exactly the same volume of crystal, the tight correlation between the two data sets suggests no major difference in crystal behavior between the two sampled volumes.

**Magnetization measurement**

The magnetization of the single crystal EuFe$_2$As$_2$ sample was measured by the vibrating sample magnetometer option of a Quantum Design Dynacool. The sample was cut into a thin octagon so that field applied along the $[1\ 0\ 0]_T$ and $[1\ 1\ 0]_T$ directions have the same (minimal) demagnetization factor. The sample was encased in GE varnish, which is known to fix the domain configuration and prevent field detwinning. In Figure 5, at T=2K the field was applied and magnetization measured along the $[1\ 0\ 0]_T$ direction (blue), i.e. at 45° to both domain easy axes, and the magnetization indicates a continuous spin canting towards a saturation at $\mu_0 H_{45°}^{sat} = 0.73$ T. Conversely, for field applied along $[1\ 1\ 0]_T$ the magnetization appears to be a combination of both a B domain spin flip at $\mu_0 H_B^{flip} = 0.45$ T and an A domain continuous spin canting which saturates at $\mu_0 H_A^{sat} = 1.15$ T, in strong agreement with the detwinned XMCD data of Figure 4. To further demonstrate this, we normalize the XMCD data of Figure 4 by the B domain 1T saturated value and average their values over the field range to simulate a 50/50 perfectly twinned sample. This data is plotted in grey in Figure 5b and is in very strong agreement with the octagonal sample, from which we conclude that the domain populations are indeed held fixed by the GE varnish.

## DATA AVAILABILITY

The authors declare that all data supporting the findings of this study are available within the article or from the corresponding author upon reasonable request.

## ACKNOWLEDGEMENTS

We thank Jannis Maiwald for useful discussion. This work was supported by NSF MRSEC at UW (DMR-1719797) and the Air Force Office of Scientific Research Young Investigator Program under Grant FA9550-17-1-0217. J.H.C. acknowledges the support of the David and Lucile Packard Foundation, the Alfred P. Sloan foundation and the State of Washington funded Clean Energy Institute. J.L. acknowledges support from the National Science Foundation under grant no. DMR-1848269. The work performed at the Advanced Photon Source was supported by the US Department of Energy, Office of Science, and Office of Basic Energy Sciences under Contract No. DE-AC02-06CH11357. JJS acknowledges the support by the U.S. Department of Energy, Office of Science, Office of Workforce Development for Teachers and Scientists, Office of Science Graduate Student Research (SCGSR) program. The SCGSR program is administered by the Oak Ridge Institute for Science and Education (ORISE) for the DOE. ORISE is managed by ORAU under contract number DE-SC0014664. All opinions expressed in this paper are the author's and do not necessarily reflect the policies and views of DOE, ORAU, or ORISE.

## AUTHOR CONTRIBUTIONS

J.J.S. conceived the project and grew the samples. G.F., Y.C, J.-W.K., P.R. and J.J.S. performed the x-ray measurements. Y.S. and P.M. performed transport and magnetization measurements on the freestanding samples. S.P. and J.L. contributed to the theoretical analysis. I.I.M. contributed to the theoretical analysis and performed DFT calculations. J.-H.C. supervised the research. J.J.S. and J.-H.C. wrote the manuscript with the input from all the authors.

**ADDITIONAL INFORMATION**

**Competing interests:** The authors declare no competing financial interests.

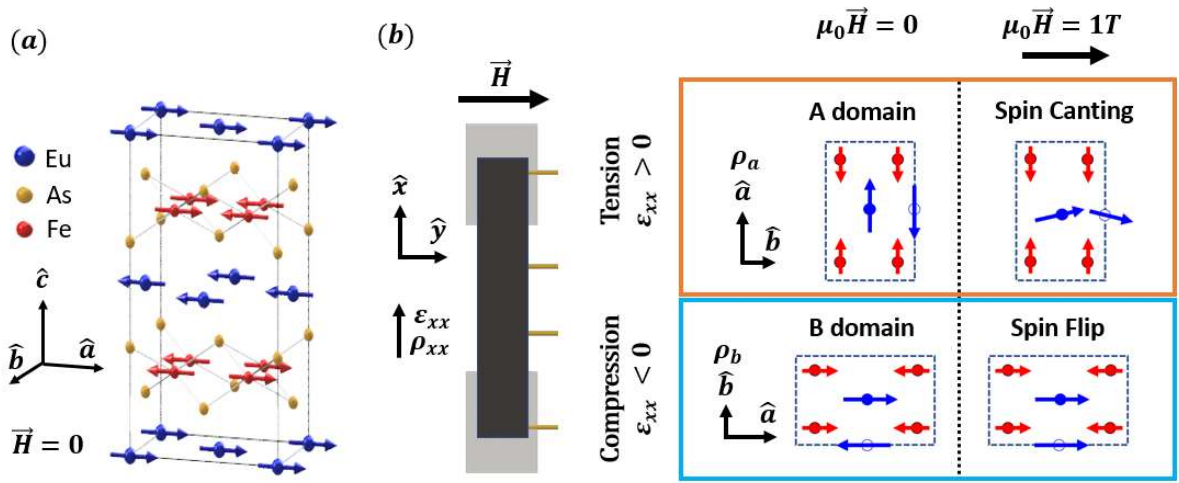

**Fig. 1** (a) EuFe$_2$As$_2$ unit cell at T=7K and zero applied magnetic field. Both Fe and Eu antiferromagnetic orders are stabilized with easy axes aligned with the longer $a$ lattice constant of the orthorhombic unit cell. (b) Uniaxial stress is applied along the $\hat{x}$ direction aligned with the orthorhombic $a/b$ orthorhombic unit cell lattice directions such that tension (compression) detwins the sample to the A (B) domain (orange/blue outline). Resistivity measurements along the stress axis measure $\rho_a$ ($\rho_b$) aligned with the $a$ ($b$) lattice constant of the A (B) domain. A magnetic field is applied perpendicular to the strain axis at 10° above parallel from the $a/b$ plane, causing a reorientation of Eu moments to align along the field direction. XMCD is proportional to the Eu magnetization along the applied field direction. For simplicity, we collapse the 4 Eu atoms and 8 Fe atoms of the doubled orthogonal unit cell into 2 Eu (blue arrows) and 4 Fe (red arrows) effective moments.

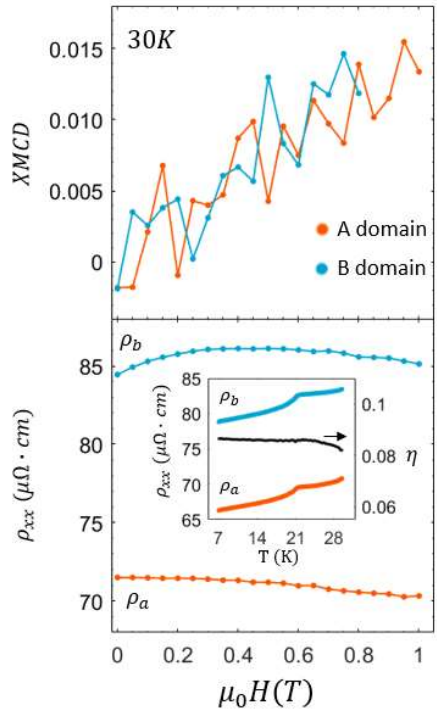

**Fig. 2** $T = 30K$ single pass applied magnetic field sweep for the detwinned A and B domains (XMCD data was not collected for B domain 0.8T-1T). XMCD above the Eu AFM ordering temperature shows a nearly isotropic response to field. $\rho_{xx}$ vs applied magnetic field reveals a minimal magnetoresistance. (Inset, left) $\rho_{xx}$ vs temperature for the detwinned A and B monodomains reveal no additional anisotropy induced at the Eu AFM ordering temperature $T_{N,Eu} = 19.1K$. (Inset, right) The resistivity anisotropy $\eta = \frac{\rho_b - \rho_a}{\rho_b + \rho_a}$ (black).

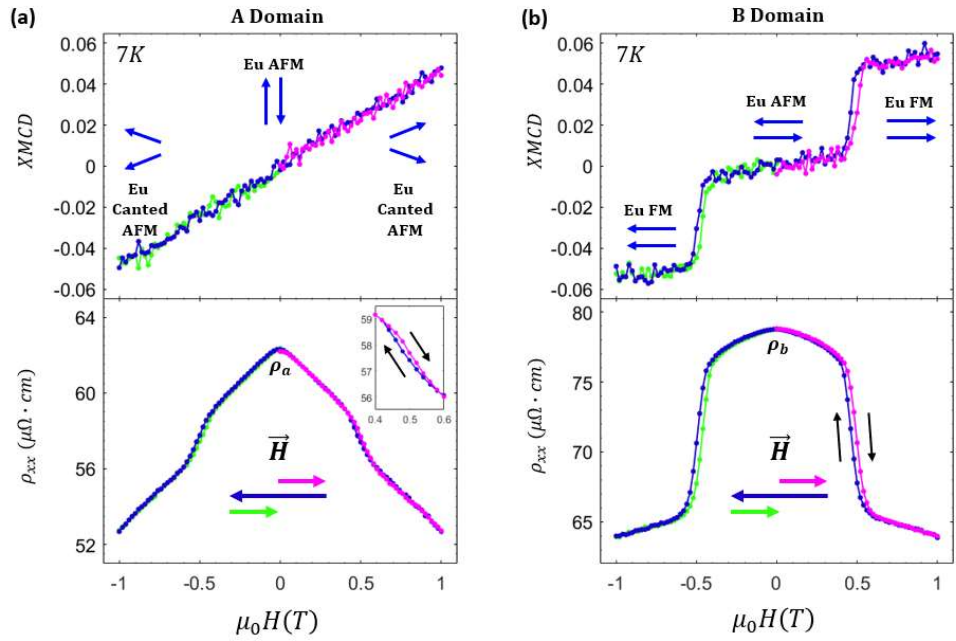

**Fig. 3** Field sweep at T=7K in the fully magnetically ordered phase presented in Fig.1a. Applied field ramped from 0T to 1T, -1T and 0T in each detwinned A (a) and B (b) monodomain states. Inset to bottom panel of (a) shows the small magnetoresistance hysteresis visible near 0.5T in the A domain.

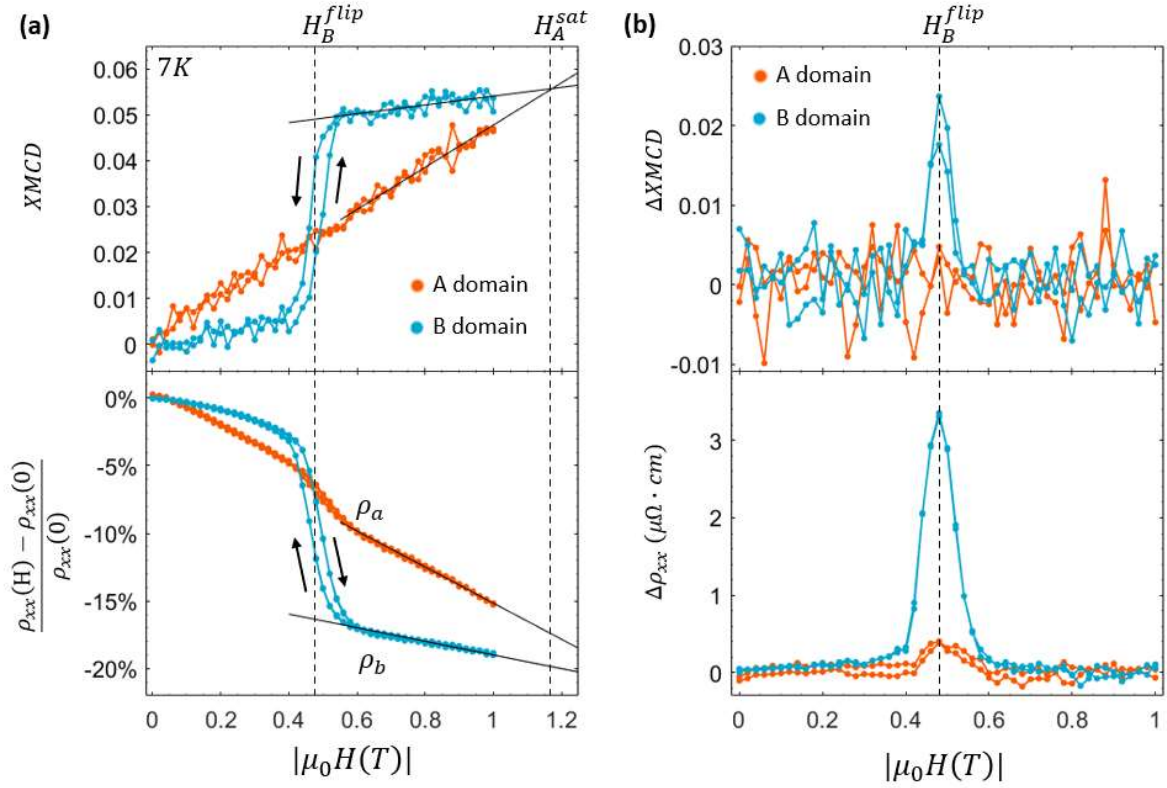

**Fig. 4** (a) Data from Fig.3a-b replotted as the average value of positive and negative field sweeps of XMCD and the magnetoresistance against the absolute value of applied magnetic field for the detwinned A (orange) and B (blue) monodomains. Linear fits (black lines) to the XMCD magnitude for $|\mu_0 H| = 0.6T$ to $1T$ indicate both values coincide at $|\mu_0 H_A^{sat}| \sim 1.17T$. The magnetoresistance for positive and negative field are nearly identical and tightly overlap. (b) The difference in XMCD and $\rho_{xx}$ for increasing and decreasing fields yields $\Delta XMCD = XMCD(H_{inc}) - XMCD(H_{dec})$ and $\Delta \rho_{xx} = \rho_{xx}(H_{inc}) - \rho_{xx}(H_{dec})$. For the B domain the peak values of $\Delta XMCD$ and $\Delta \rho_{xx}$ coincide at $\mu_0 H_B^{flip} = 0.48T$.

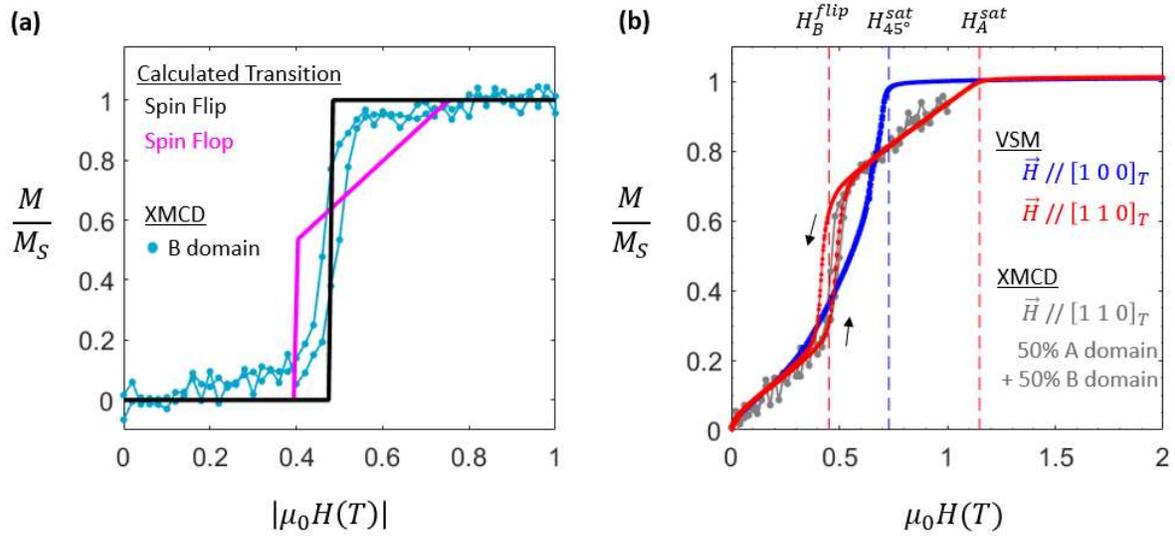

**Fig. 5 (a)** The XMCD data of Fig.4a for the B domain normalized to the 1T mean value. Black (magenta) line represents the T=0K metamagnetic spin flip (spin flop) transition calculated using the anisotropic JKW (isotropic JK) model (see main text). **(b)** Octagon sample magnetization with field applied along the tetragonal $[1\,0\,0]_T$ (blue) and $[1\,1\,0]_T$ (red) directions. Critical fields marked by dashed lines. Magnetization normalized by 1T saturation value along $[1\,0\,0]_T$. (grey) Average of the two domain XMCD data of Figure 4a, normalized by the B domain saturated value at 1T.